# Evidence of interfacial asymmetric spin scattering at ferromagnet-Pt interfaces

Van Tuong Pham, Maxen Cosset-Chéneau, Ariel Brenac, Olivier Boulle, Alain Marty, Jean-Philippe Attané[*], Laurent Vila[*]

University Grenoble Alpes, CEA, CNRS, Spintec, F-38000 Grenoble, France

[*] jean-philippe.attane@cea.fr and laurent.vila_@cea.fr

ABSTRACT

We measure the spin-charge interconversion by the spin Hall effect in ferromagnetic/Pt nanodevices. The extracted effective spin Hall angles (SHAs) of Pt evolve drastically with the ferromagnetic (FM) materials (CoFe, Co, and NiFe), when assuming transparent interfaces and a bulk origin of the spin injection/detection by the FM elements. By carefully measuring the interface resistance, we show that it is quite large and cannot be neglected. We then evidence that the spin injection/detection at the FM/Pt interfaces are dominated by the spin polarization of the interfaces. We show that interfacial asymmetric spin scattering becomes the driving mechanism of the spin injection in our samples.

Spin-charge interconversions (SCICs) which are pioneered by spin Hall effect (SHE) in heavy metals[1] have emerged as alternative ways for spin manipulation and are at the core of spinorbitronics[2]. The efficiency of the charge to spin conversions is a crucial issue, as it determines how much torque that can be generated to write information by the spin-orbit torques (SOTs)[3,4]. It is also essential in the reciprocal mechanism, for example in the reading process[5] of magneto-electric spin–orbit logic devices[6], a beyond-CMOS technology where a spin current emitted by a ferromagnetic (FM) element generates a transverse charge current, whose sign depends on the magnetic states. Beyond the conversion rate by the spin-orbit coupling (SOC), the interface is the other key element to be optimized. Indeed, the spin memory loss[7,8], the disorder at the interface[8,9], the spin reflection/spin backflow [10,11] due to the spin impedance mismatch and interfacial-SOC-based spin-transparency[12] have been identified as key factors in the process of SCIC in simple FM/SOC bilayers. The spin memory loss decreases the spin current effectively injected, while the disorder at the interface can actually lead to high interface resistance[8]. Together with the relaxation by the injecting element, they are both detrimental for spin injection. For instance, a high spin transmission increases the spin current crossing the interface but also the spin back flow.

In this Letter, direct and inverse SHE are electrically measured in FM/Pt nanostructures made of different FM materials[13,14]. We find that the amplitude of SHE signals depends drastically on the magnetic materials ($Co_{60}Fe_{40}$, Co, and $Ni_{81}Fe_{19}$). Using finite element method (FEM) simulations and assuming the interface to be transparent, we extract effective spin Hall angles (SHAs) varying strongly with the FMs, with high values in line with spin torque-FM resonance (-FMR) experiments[10,15]. The values of the interface resistances are then carefully measured for each FM/Pt interface. This measured spin-independent additional resistance should lead to the decrease of the spin signals, which is inconsistent with our observations. We show that the interfacial scattering asymmetry, which is a known parameter of the spin transport in GMR stacks[16], has thus to be taken into account for the estimation of the SCIC. Although it is usually overlooked in spinorbitronics, it allows completing the spin transport picture, with a major role in our metallic FM/SOC nanostructures. Contrarily to what could be expected, resistive FM/Pt interfaces can favorably enhance the spin Hall signal, which is useful for the spin-orbit magnetic state readout[5,6].

Figures 1a, 1b and 1c illustrate our principle and measurement configuration of direct SHE[13,14]. As sketched in Fig. 1a, a charge current density ($J_C$) flows along a nanowire made of platinum, a strong SOC material. The direct SHE converts the applied charge current density into a transverse spin current density (Js), leading to two opposite spin accumulations (SA) localized at the top and bottom surfaces of the Pt wire. The charge to spin current conversion is defined as $J_S = \theta_{SH} J_C$, where $\theta_{SH}$ is the SHA of SOC material. The associated electrochemical potential landscape is sketched in the right of Fig. 1a. The SA on the top surface of the SOC wire is detected using FM electrodes by aligning their Fermi level with the electrochemical potential ($\mu_\uparrow$ or $\mu_\downarrow$) of the SA [Fig. 1b-i]. The voltage between the two electrodes thus depends on their magnetic configurations (parallel or anti-parallel). A resistive FM/Pt interface can contribute to probe the SA [Fig. 1b-ii]. An additional electrochemical potential appears at the interface due the continuity of the spin current through the resistive interface.

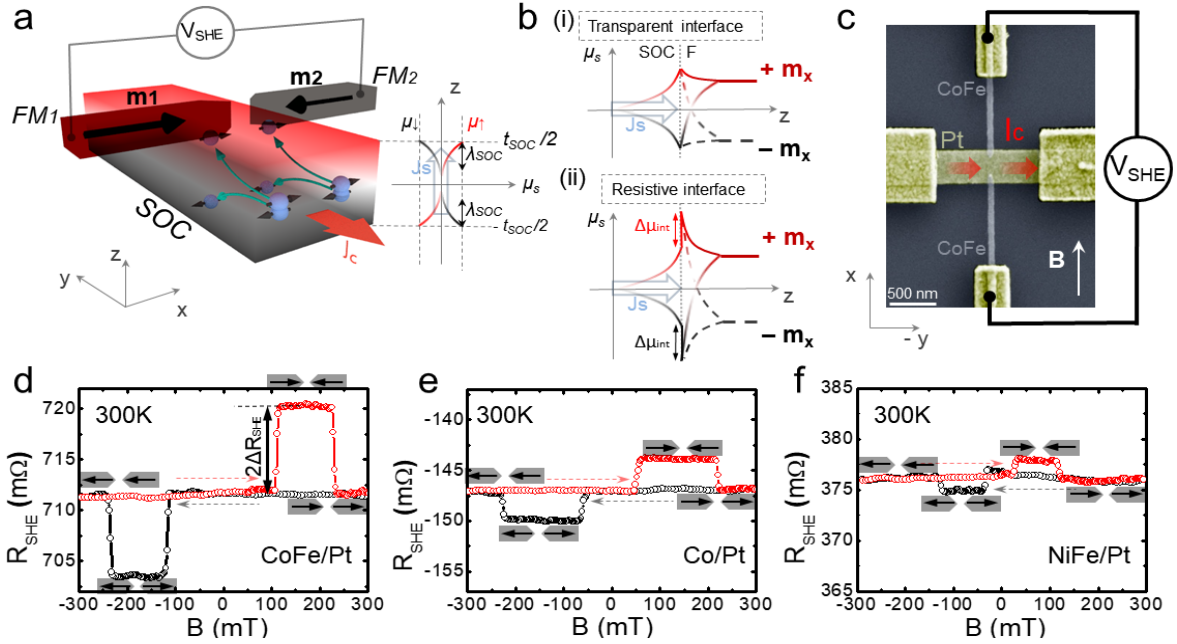

**Fig. 1.** (a) Sketch of direct SHE measurement using the FM/SOC nanostructure. Left: Two FM electrodes probe the spin accumulation (SA) induced at the top surface of a SOC wire by the direct SHE. The red volume is spin-up rich (+x polarization) while the grey one is spin-down rich (-x polarization). The black arrows indicate the vector magnetizations of the electrodes (+$m_x$/-$m_x$). Right: electrochemical potential landscape within the SOC material. (b) Electrochemical potential landscape at one FM/SOC interface of the device, in the case of a transparent interface (i), and in the case of a resistive interface with an interfacial spin asymmetry (ii). The red and black curves correspond to the electrochemical potential of electrons whose spins are polarized along x and –x, respectively. In (ii), $\Delta\mu_{int}$ is an additional electrochemical potential due to the resistive interface contribute to spin detections. (c) Scanning electron microscopy (SEM) image of a CoFe/Pt sample with a sketch of direct SHE measurement configuration. (d), (e), and (f) are the experimental field sweep loops measured along x on CoFe/Pt, Co/Pt, and NiFe/Pt samples, respectively, with identical geometrical parameters: nanowire widths of $w_{Pt} = 400\ nm$ and $w_{FM} = 50\ nm$ and thicknesses $t_{Pt} = 7\ nm$ and $t_{FM} = 15\ nm$. The black arrows indicate the configuration of the magnetic electrodes when sweeping the magnetic field.

When the two FMs have parallel magnetizations, the same electrochemical potential is probed, and the voltage between the two FMs is zero. In the anti-parallel states of the magnetizations, the two FMs are probing opposite electrochemical potentials, and a non-zero voltage is measured, with opposite signs for the head-to-head and tail-to-tail magnetic configurations[13,14]. The spin Hall signal, in Ohm, is the measured voltage divided by the applied current. The amplitude of the SHE signals is the difference of the measured resistance between the anti-parallel and parallel magnetic states of the FMs.

As expected, asymmetric spin signals are observed as shown in Figs. 1d, 1e, and 1f for CoFe/Pt, Co/Pt, and NiFe/Pt systems, respectively. The signal amplitudes are found to vary with the FM material, with values for CoFe/Pt, Co/Pt, and Py/Pt of 8.0 ± 0.5, 4.1 ± 0.5, and 1.7 ± 0.4 mΩ, respectively. Note that in these experiments, we can measure either the direct SHE or the inverse SHE, and that they both give exactly the same signal amplitude[13,14]. Note also that the slight contribution of the anomalous Hall effect due to the ferromagnetic electrodes to the spin signal can be neglected [SM S4]. Additionally, second harmonic measurement allows disregarding thermal effect contributions [SM S6].

The efficiency of the SCIC in Pt is related to the product of the SHA by the spin diffusion length, $\theta_{SH}\lambda_{Pt}$[17]. Despite the fact that the SHA depends on the Pt resistivity, the product $\theta_{SH}\lambda_{Pt}$ is expected to be independent of the resistivity, because of the intrinsic origin of the SHE in Pt at room temperature[18,19]. Using spin pumping-FMR techniques, $\theta_{SH}\lambda_{Pt}$ values in the range of 0.17 to 0.24 nm are usually extracted[7,18,20]. In particular we have extracted a $\theta_{SH}\lambda_{Pt} = 0.19\ nm$[7]. These values of $\theta_{SH}\lambda_{Pt}$ are in line with first-principles calculations of the Pt spin Hall conductivity[21,22]. If the Elliott-Yafet mechanism[23] is dominant, the spin diffusion length can be estimated from the resistivity value ($\rho_{Pt}$). For our Pt wires, $\lambda_{Pt}$ = 2.6 nm and $\theta_{SH} = 0.073$ are calculated from the measured $\rho_{Pt}$ = 28.0 μΩcm.

Using FEM simulations in the case of transparent interface (Ref. 13) and the transport parameters we have previously estimated in our nanostructures (Ref. 24), we extract quite different values of effective $\theta_{SH}\lambda_{Pt}$ for the different FMs, as shown in the column 2 of *Table 1*. Although we take into account the variations of the spin polarizations and the other transport parameters in FMs, we do not obtain a consistent value of $\theta_{SH}\lambda_{Pt}$. The corresponding effective SHA of Pt would be of 0.19, 0.34 and 0.10 for CoFe, Co and NiFe electrodes respectively. These scattered values imply that this analysis fails to describe the SCIC in the systems.

In order to clarify the role of the interface in the spin signals, we measured the FM/Pt interface resistances. The samples consist of simple cross made of a thick FM wire (CoFe, Co, and NiFe), patterned transversally on top of a Pt wire (Fig. 2a). A current is applied from a Pt branch to a FM branch, so that it is flowed through the interface. The voltage drop is probed in-between the two other branches, which sum up the voltage drops due to the volumes and the interface. We used 3D FEM modeling to take into account the contribution of the bulk resistances so that the interface resistance can be estimated[25]. The interface specific resistance is defined by the product of its area and resistance, RA, and is varied in the FEM. As shown in Fig. 2b, the current line crossing the interface for a large RA are more homogeneous while for lower RA values, the current lines are more complex. Experimental and simulated values are plotted in Figs. 2c-2e for CoFe, Co and Py, respectively.

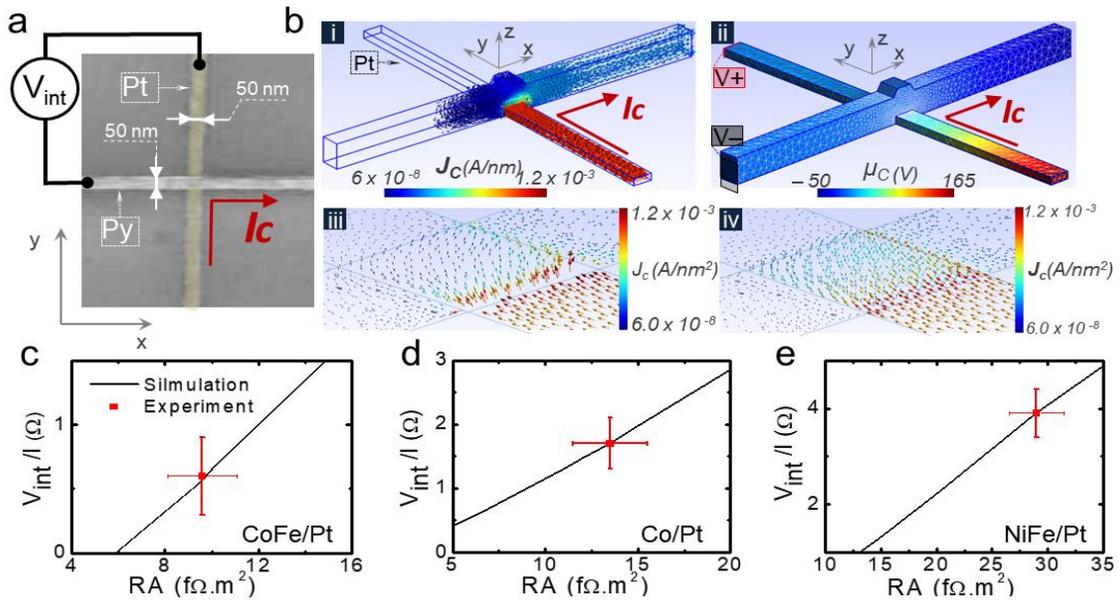

**Fig. 2.** (a) SEM image of a NiFe/Pt cross, and sketch of the electrical setup for interface resistance measurements. (b) 3D-FEM simulation results of charge transport through the interface: (i) and (ii) are the distributions of the charge current density ($J_C$) and potential ($\mu_C$) landscape, (iii) and (iv) show the charge current distribution in the interface plane, for a transparent interface (iii) and an interface resistance of 29 f$\Omega$m$^2$ (iv). (c), (d), and (e) are the expected resistance as a function of the interfacial RA of CoFe/Pt, Co/Pt, and NiFe/Pt systems, respectively. From the simulation curves we deduced the interface RA from the measured $V_{int}/I$ value. The measured $V_{int}/I$ value is an average of measurements obtained on 3 to 5 different nanostructures.

The obtained values of RA products [Table 1] are quite large, in the range of 10-30 f$\Omega$m², and more than one order of magnitude larger than that of FM/Cu or Pt/Cu[25]. A possible contribution to these high RA values is the additional scattering due to the disorder at the interface[26,27,28]. For instance, the measured RA value of the NiFe/Pt interface agrees well with that calculated by Liu *et al.*[8] which shows that it is proportional to $\rho_{Pt}$. Correspondingly, the measured RAs of the Co/Pt and CoFe/Pt interfaces are about one order of magnitude larger than those measured in systems of much lower resistivities[28,29,30].

We implemented the role of the interface resistance in the SHE measurements in the 3D FEM simulations (cf. Refs. 13, 14, and 25 for the method]. The two quantities, the measured RA and spin scattering asymmetry γ, are introduced to the boundary condition for the spin transport at the interface as shown in SM S5. Particularly, in the framework of the two-current model, the interface resistance is introduced as a potential barrier with different heights depending on the spin polarization of the electron. The interfacial resistance is then given by $RA_\uparrow$ for the spin up electrons, and $RA_\downarrow$ for the spin down electrons. This generates an electrical resistance $RA = 1/(RA_\uparrow^{-1} + RA_\downarrow^{-1})$ where the spin scattering asymmetry, $\gamma = (RA_\uparrow^{-1} - RA_\downarrow^{-1})/(RA_\uparrow^{-1} + RA_\downarrow^{-1})$, acts as a source of spin polarization.

Table 1. *Parameters and retrieved values in the different FM/Pt systems.* $p_F$: *polarization of the bulk FMs estimated using the spin absorption technique*[24]. $\theta_{SH}\lambda_{Pt}$ *(trans.): effective $\theta_{Pt}\lambda_{Pt}$ estimated by FEM simulation when assuming transparent interfaces.* **RA**: *interfacial RA estimated from the experiments [Fig. 2].* $\theta_{SH}\lambda_{Pt}$ *(RA): extracted values of effective $\theta_{SH}\lambda_{Pt}$ when the measured RA and γ = 0 are introduced in the FEM model.* **γ**: *Extracted value of the interfacial spin scattering asymmetry calculated with the measured RAs and assuming that the intrinsic value of $\theta_{SH}\lambda_{Pt}$= 0.19 nm.*

| Devices | $p_F$ | $\theta_{SH}\lambda_{Pt}$ (*trans.*) (nm) | RA (f$\Omega$m$^2$) | $\theta_{SH}\lambda_{Pt}$ (*RA*) (nm) | γ |
|---|---|---|---|---|---|
| CoFe/Pt | 0.48[24] | 0.60[13,14] | 9.6 ± 1.5 | 1.83 ± 0.10 | 0.46 ± 0.05 |
| Co/Pt | 0.17[24] | 0.81 | 13.5 ± 2.0 | 2.38 ± 0.10 | 0.17 ± 0.03 |
| Py/Pt | 0.22[24] | 0.22[13,14] | 29.0 ± 2.5 | 0.93 ± 0.10 | 0.070 ± 0.015 |

Figure 3b plots the amplitude of the simulated spin signal as a function of RA, with and without spin scattering asymmetry for CoFe/Pt system. If parameter γ is set to zero, the spin signal drastically decreases as the interface resistance increases (green line in Fig. 3b). Indeed, the electron transit time through the interface becomes larger than the spin relaxation time in FM, and thus the SA depolarizes[31]. To fit the experimental data, this would require increasing the $\theta_{SH}\lambda_{Pt}$ value by roughly a factor of 10 with respect to the transparent interface case. Using this assumption of a non-polarized interface resistance, we obtain $\theta_{SH}\lambda_{Pt}$ values of the order of 1.83 nm, or equivalently a SHA of 0.74 for a $\lambda_{Pt}$ of 2.6 nm according to room temperature Pt resistivity ($\rho_{Pt} = 28\ \mu\Omega cm$)[24]. This effective value of SHA is in line with a recent report[10] which technically underestimate spin transmission when simply taking into account the spin backflow due to the spin resistance mismatches. The SHA can be even unrealistic for the case of Co/Pt ($\theta_{SH} \approx 1$). Furthermore, the extracted values are different in the three systems (Table 1).

The Rashba coupling at the FM/Pt interface could contribute to the spin signals. However, metallic systems require a high conducting interface to confine the interfacial electrical current for efficient SCIC by Rashba coupling[32]. Our systems are ex-situ deposited and a strong SOC at interface is unlikely. Additionally, the interfacial SCIC is rather small in comparison to bulk effect[33, 34].

The spin resistances determine the strength of spin injection/detection by a ferromagnet. The measured RAs are in the 10-30 f$\Omega$.m$^2$ range, whereas the FM spin resistances are of the order of a few f$\Omega$.m$^2$ (1.98, 1.61, and 1.63 f$\Omega$.m$^2$ for Co, CoFe, and NiFe, respectively). It turns out that the injected spin current is mainly determined by the interface contribution, and thus by the spin asymmetry parameter. We argue in the following that this key parameter of the spin-dependent transport has to be taken into account. Indeed, when assuming $\theta_{SH}\lambda_{Pt} = 0.19$ nm, we recover a $\gamma$ value to be equal to the polarization of the bulk FM $p_F$, using the experimentally measured RA of CoFe/Pt (9.7 f$\Omega$m$^2$).

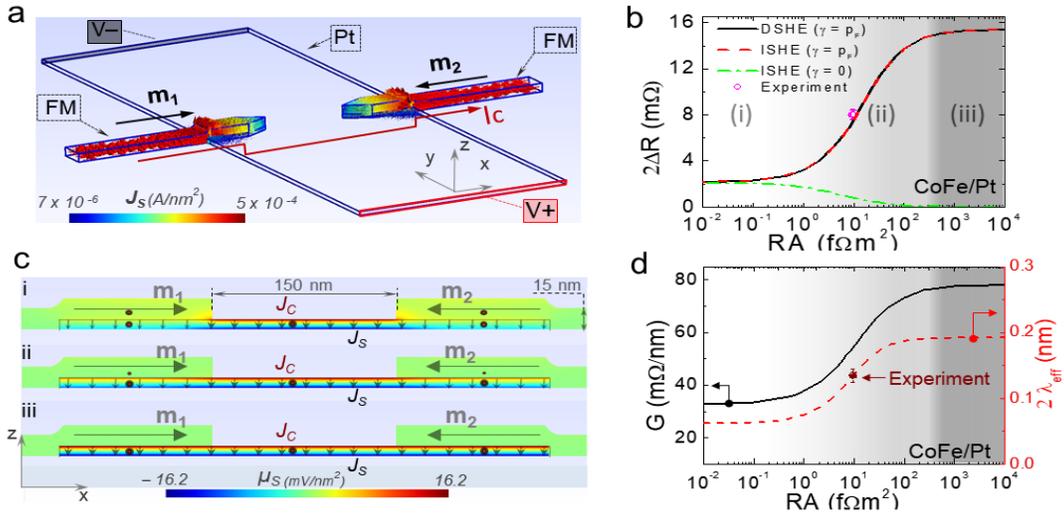

**Fig. 3.** (a) Example of FEM simulation for the SCIC device in the inverse SHE configuration with the geometrical parameters of Fig. 1. A charge current $I_C$ is injected in between the ferromagnetic electrodes, which possess antiparallel magnetizations. The inverse SHE converts the spin current flowing in the z-direction into a charge current along the Pt wire (y-direction). The SA profile is presented in S.M. S1. (b) Spin signal amplitudes as a function of the RA for CoFe/Pt. The simulation indicates three regions corresponding to three regimes of the interface: transparent (i), intermediate (ii), and resistive (iii). (c) x-z cross section at the sample centre of the electro-chemical potential distribution (proportional to the spin density). It presents the SA profile in the direct SHE configuration (Fig. 1a), for values of RA corresponding to three cases (i, ii, and iii). The applied charge current density ($\mathbf{J_C}$) is in the –y direction (red-black dots in different size shows the different local $\mathbf{J_C}$ due to the electrical shunting by FM). The SHE creates a spin current in the z-direction ($\mathbf{J_S}$) (small black arrows in Pt, the different size correspond to different produced local $\mathbf{J_S}$). $\mathbf{m_1}$ and $\mathbf{m_2}$ are the magnetizations of FMs. (d) Plots of the geometry factor (G) (see more SM S4 and ref. 5) and of the effective spin-charge conversion rate ($\lambda_{eff}$) as a function of RA, for $\gamma = p_F$, *i.e.*, the spin polarization of FM electrode is kept at the interface.

Obtaining high values of direct SHE and inverse SHE signals are necessary, for instance for developing SO magnetic readout[5]. As the interface resistance is basically a resistance to the spin current injection, one can intuitively expect that the signals will decrease when increasing the interface resistance. If this is indeed the case when considering that $\gamma = 0$, our results show that for $\gamma \neq 0$, increasing the interface resistance should actually lead to higher signals. Fig. 3c supplies a guideline to enhance the inverse SHE signal by increasing the interfacial RA, and Fig. 3d shows the mechanism behind it. The inverse SHE signal can be expressed as an electromotive strength of a current source, as it depends on the internal resistance of the structure and on the inverse SHE efficiency. The spin signal amplitude can be simply written $\Delta R = 2G\lambda_{eff}$, where G is the geometry/resistance factor, and $\lambda_{eff}$ is the effective efficiency of the spin to charge conversion which includes spin injection and spin to charge current conversion efficiencies. Fig. 3d shows the change of the effective conversion rate ($\lambda_{eff}$) as a function of RA, calculated using the 3D FEM model (SM S3). When assuming that $\gamma = p_F$, both G and $\lambda_{eff}$ are proportional to the value of the interface resistance (RA) and favors high spin signal amplitudes [ref. 5].

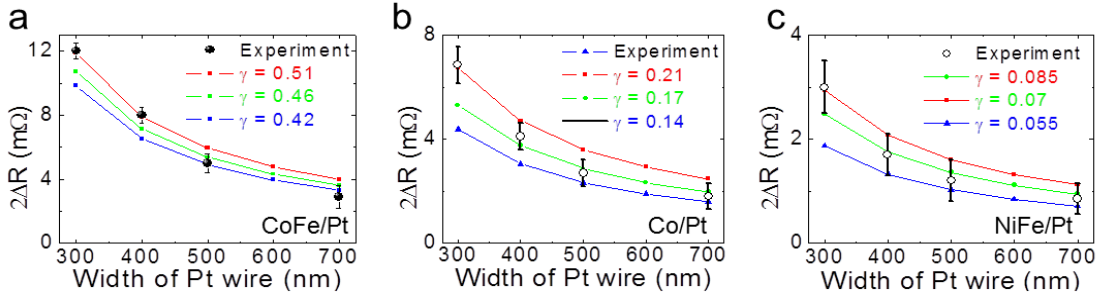

**Fig. 4.** Simulated and experimental values of the amplitude of SHE signals vs. the Pt wire width in CoFe/Pt (a), Co/Pt (b), and NiFe/Pt (c) systems. The simulations are performed assuming the same Pt SHE efficiency ($\theta_{Pt}\lambda_{Pt} = 0.19$ nm). $t_{FM}$, $t_{Pt}$, and $w_{FM}$ are the same as in Figs. 1(c), 1(d), and 1(f).

Figures 4a-4c plot the amplitude of the simulated and experimental spin signals, as a function of the Pt wire width, in the three systems[13,14]. Here, we keep assuming that $\theta_{Pt}\lambda_{Pt}$ is of the order of 0.19 nm. The single free parameter, $\gamma$, that allows this consistent extraction of $\theta_{SH}\lambda_{Pt}$ is then found to be 0.46, 0.17 and 0.07 for CoFe/Pt, Co/Pt and Py/Pt interfaces. The results match quite well with the polarization of the bulk ferromagnets[29], except for the case of NiFe/Pt interface which is a footprint of the high spin memory loss at this interface. Particularly, if the spin flip ratio of $\delta = 3.8$ from ref. 8 is introduced, the effective spin resistance of the interface is decreased and consequently $\gamma$ becomes similar to the value of $p_F$ of bulk NiFe. A simulation has been made to verify this agreement as shown in Fig. S6 in SM. Note that a factor 3 between the bulk polarization and $\gamma$ has been theoretically predicted by K. Gupta *et al.* 9, which has taken into account the magnetic disorder at the NiFe/Pt interface. Our results tend to prove that large interface resistances and non-zero $\gamma$ values are at the origin of the large and scattered values of the effective SHA, usually extracted assuming either completely transparent interfaces or resistive interfaces without taking into account $\gamma$.

To conclude, we studied the importance of the FM/Pt interface in nanostructures for SHE measurements, with different FM materials. We show that the measured interface resistance values are relatively high with respect to the FM spin resistances. This leads to high signals, and when the interface polarization is not taken into account, to an overestimation of the SHA with a value that depends on the injecting FM material. We thus argue that in addition to the interfacial resistance a spin scattering asymmetry has to be introduced. Co and CoFe have a high spin scattering asymmetry ($\gamma \approx p_F$) at the interface with Pt, favorable for obtaining large spin signals, while the NiFe/Pt has a small effective spin scattering asymmetry (or presents a high value of the interfacial spin flip ratio) leading to a lower spin injection. These results emphasize that the interfacial resistance and magnetic properties need to be carefully taken into accounts in the FM/SOC bilayers to estimate the SHE efficiency of SOC materials.

As outlooks for spintronic applications, we anticipate that the interface resistance can help improving the efficiency of injecting/detecting the spin current, *i.e.* producing better sources of the electromotive forces. Particularly, reaching high resistances of the magnetic interfaces while conserving high γ values would allow enhancing the spin signals for the SO magnetic readout[5] and the current supply in MESO logic[6], for instance using oxide interfaces[35] and topological-insulators based systems[36]. SOTs[3,4] require transparent interfaces. Nonetheless, too transparent interfaces can increase the spin back flow.

**Acknowledgements**

The devices were fabricated in the Platforme Technologie Amont in Grenoble. We acknowledge the support from the labex laboratory LANEF of Univ. Grenoble Alpes and the ANR Project TOPRISE.